\documentclass[trackchanges]{aastex701}

\usepackage{amsmath}
\usepackage{mathtools}

\newcommand{\isot}[2]{$^{#2}\mathrm{#1}$}
\newcommand{\isotm}[2]{{}^{#2}\mathrm{#1}}
\newcommand{\Xisot}[2]{$X(\isotm{#1}{#2})$}
\newcommand{\Xisotm}[2]{X(\isotm{#1}{#2})}
\newcommand{\Rconvm}{R_{\mathrm{conv}}}
\newcommand{\Rconv}{$\Rconvm$}
\newcommand{\Mconvm}{M_{\mathrm{conv}}}
\newcommand{\Mconv}{$\Mconvm$}
\newcommand{\Msun}{\ensuremath{\mathrm{M}_\odot}}

\newcommand{\vrmsm}{U_{\mathrm{rms}}}
\newcommand{\vrmsrm}{\vrmsm (r)}
\newcommand{\vrms}{$\vrmsm$}
\newcommand{\vrmsr}{$\vrmsrm$}
\newcommand{\epsnucm}{\dot{\epsilon}_\mathrm{nuc}}
\newcommand{\epsnuc}{$\epsnucm$}
\newcommand{\epsnum}{\dot{\epsilon}_\mathrm{\nu}}
\newcommand{\epsnu}{$\epsnum$}
\newcommand{\epsthermm}{\dot{\epsilon}_\mathrm{therm}}
\newcommand{\epstherm}{$\epsthermm$}

\newcommand{\unitstyle}{\mathrm}
\newcommand{\second}{\unitstyle{s}}   

\newcommand{\K}{\unitstyle{K}}
\newcommand{\erg}{\unitstyle{erg}}        
\newcommand{\gcc}{\unitstyle{g~cm^{-3}}} 
\newcommand{\kms}{\unitstyle{km~s^{-1}}} 

\newcommand{\maestro}{{\sffamily MAESTROeX}}

\newcommand{\microphysics}{{\sffamily Microphysics}}
\newcommand{\pynucastro}{{\sffamily pynucastro}}
\newcommand{\amrex}{{\sffamily AMReX}}

\newcommand{\dif}{\partial}
\begin{document}
\title{On the Importance of the Convective Urca Process in 3D Simulations of a Simmering White Dwarf}
\author[0009-0000-1964-7734]{Ferran Poca-Amor\'{o}s}
\affiliation{Universitat Polit\`{e}cnica de Catalunya, Barcelona, Spain}
\email{ferran.poca@estudiantat.upc.edu}

\author[0000-0002-5419-9751]{Brendan Boyd}
\affiliation{Department of Physics and Astronomy, 
Stony Brook University, Stony Brook, NY 11794-3800, USA}
\affiliation{Institute for Advanced Computational Science,
Stony Brook University, Stony Brook, NY 11794-5250, USA}
\email{boyd.brendan@stonybrook.edu}

\author[0000-0002-9538-5948]{Dean M.\ Townsley}
\affiliation{Department of Physics and Astronomy, University of Alabama, Tuscaloosa, AL 35487-0324, USA}
\email{Dean.M.Townsley@ua.edu}

\author[0000-0001-5525-089X]{Alan Calder}
\affiliation{Department of Physics and Astronomy, 
Stony Brook University, Stony Brook, NY 11794-3800, USA}
\affiliation{Institute for Advanced Computational Science,
Stony Brook University, Stony Brook, NY 11794-5250, USA}
\email{alan.calder@stonybrook.edu}

\correspondingauthor{Ferran Poca-Amor\'{o}s}
\email{ferran.poca@estudiantat.upc.edu}

\begin{abstract}
Type Ia supernovae are bright thermonuclear explosions that are important to numerous areas of astronomy. However, the origins of these events are poorly understood. One proposed setting is that of a near Chandrasekhar mass white dwarf that undergoes runaway carbon burning in the core. During the thousand years leading up to the explosion, the white dwarf undergoes a simmering phase where slow carbon burning heats the core and drives convection. A poorly understood aspect of this phase is the convective Urca process, which links convection with weak nuclear reactions. We use the low Mach number code \maestro\ to perform full 3D simulations as is required to accurately capture the turbulent convection. We present simulations with and without the A=23 convective Urca process, which have relaxed to a steady state. We characterize the effects of the convective Urca process on the neutrino losses, the nuclear energy generation,  and the convective boundary.  We find that the size of the convection zone is substantially reduced by the convective Urca process, though convection still extends past the Urca shell. Our findings on the structure of the convective zone and the compositional changes can be used to inform 1D stellar models that track the longer-timescale evolution.
\end{abstract}


\section{Introduction} \label{sec:intro}

Type Ia Supernovae (SNe Ia) are widely understood as thermonuclear
explosions of white dwarfs (WDs) in binary systems. 
Most SNe Ia events ($\sim 70\%$) follow the so called ``Philips relation" \citep{phillips1993} and accordingly may be calibrated, making SNe Ia ideal standard candles. 
This feature has made them a key tool in cosmology research \citep{riess1998, perlmutter1999}. Despite the widespread use of SNe Ia as cosmological indicators, plenty of characteristics of these explosions still remain enigmatic, for example the exact progenitor
system that leads to SNe Ia. Numerous models have been studied for the progenitor system. These can be broadly categorized as scenarios where the WD's companion star is another WD (double-degenerate) or a non-degenerate star (single-degenerate), and/or where the WD explodes close to or well below the Chandrasekhar-mass limit (Chandrasekhar-mass or sub-Chandrasekhar-mass progenitors) \citep{liu2023}. In this paper, we focus on the Chandrasekhar-mass model, where a carbon-oxygen WD accretes material from a non-degenerate companion
star growing up to near the Chandrasekhar-mass limit before thermonuclear runaway ignites in the core \citep{nomoto1984,woosley1986}.

In this scenario, the central temperature and density of the WD increases during the accretion process, reaching the conditions for central carbon burning. This
begins a simmering phase in which carbon burning drives subsonic core convection
for about 1,000 to 10,000 years before the explosion \citep{woosley2004}. As the central temperature increases, the convection zone expands and the burning rate increases to the point where the carbon burning timescale is shorter than the convection timescale. At this point, a flame will ignite near the WD center and incinerate the WD partially or completely. The exact mechanism that will drive the explosion is still a matter of active discussion \citep{calderetal2013, hillebrandt2013, liu2023}. 

The carbon burning during the simmering phase alters the WD composition. In particular, through $\isotm{C}{12}(\isotm{C}{12},p)\isotm{Na}{23}$ and subsequent electron capture reactions, the carbon burning process generates neutron-rich material, reducing the electron fraction \citep{piro2008, chamulak2008}. As a consequence of these compositional changes, some of the observable features of the SNe Ia can be affected, such as the $\isotm{Ni}{56}$ and other Fe-group nuclei production \citep{Umeda1999, timmes2003}. In order to track the composition of the WD during the full timescale of the simmering phase, one-dimensional stellar evolution models have been used \citep{martinez-rodriguez2016, piersanti2017, schwab2017a, piersanti2022}. However, due to the limited understanding on how to treat convective boundaries, modeling the convective mixing in the simmering phase has become a particularly challenging problem. Realistic modeling of the convective mixing requires considering the effects of the convective Urca process, which is a linking between weak nuclear reactions and convection. 

Originally proposed in \citet{paczynski1972}, the convective Urca process leaks energy away from a degenerate star via the neutrino emission produced in $\beta$-decay and electron capture reactions. The process occurs because the convection mixes the material from dense regions, where electron captures can occur, to less dense regions, where $\beta$-decays occur. As the material is cyclically mixed from high density to low density and back, there is continuous neutrino emission. Further study of the convective Urca process has shown that, even though it can result in some local cooling of the star, it cannot prevent the thermonuclear runaway \citep{bruenn1973, couch1975}. Instead, it has been proposed as a process that can constrain the growth of the convective core during the simmering phase \citep{stein-wheeler2006, piersanti2022}. However, the exact impact of the convective Urca process in the simmering phase evolution and in the features of the convective core is still incompletely understood.

Results published in \cite{lesaffre2005} showed,  using two-stream analytical approach, that the convective Urca process may hinder/limit the strength of convection. Later, the convective Urca process was further studied in \citet{stein-wheeler2006}, with 2D simulations using an implicit hydrodynamic code with boosted
reaction rates and a wedge geometry. Similarly to the analytical approach, they found that the convective Urca process limits the size of the convective core to the radius of the so-called Urca shell (region where $\beta$-decay and electron capture reaction rates are equal). However, these results cannot be considered conclusive, as convection is an inherently turbulent 3D problem, and the work presented in \citet{stein-wheeler2006} relies on the 2D treatment of convection and a wedge geometry which excludes the center of the WD. In fact, more recent full 3D studies published in \citet{boydetal2025} (which will be referred to as B2025) demonstrate that, in a late stage of the simmering phase, the convective core can extend considerably further than the Urca shell. In addition, they showed that when using a 3D model for convection, the mixing had very strong directional dependencies, supporting the need for further modeling of the 3D convection. Understanding the role of the convective Urca process can be very relevant to the use of 1D stellar evolution models because the evolution of a WD in the simmering phase can be very sensitive to different prescriptions of this process \citep{denissenkov2015, piersanti2022}.

In this paper, following the work started in \citet{willcox2018} and more recently continued in B2025, we present full 3D hydrodynamic simulations of a WD in the simmering phase using the \maestro\ code \citep{fan2019}. In particular, to characterize the effects of the convective Urca process in the convective core, we run 3D simulations with and without considering such process. We directly compare two of these simulations which have relaxed to steady-states. And by doing so, we are able to characterize the effects that the convective Urca process has on the convective core size, characteristic velocities and energy generation.

In Section \ref{sec:2}, we discuss the convective Urca process in additional detail. Section \ref{sec:3} briefly describes \maestro\ equations and numerical methodology. Later, in Section \ref{sec:4}, we introduce the initial models and the nuclear networks used in the simulations. In Section \ref{sec:5}, we present and analyze the final steady-state of the simulations we ran, including results about the convective motion, energy generation, buoyancy and carbon burning. Then, in Section \ref{sec:6}, we discuss the impact of the convective Urca process in the results obtained through the simulations, and briefly note the limitations of this study. Lastly, in Section \ref{sec:7}, we draw our conclusions and point to future work.

\section{The Convective Urca Process}
\label{sec:2}
The Urca process is the combination of a $\beta$-decay and an electron capture
reactions that connect a pair of nuclei, called an Urca pair. The
most relevant Urca pair in the context of the simmering phase of a WD
is the $A=23$ Urca pair,  \isot{Na}{23} -- \isot{Ne}{23}, linked by 
\begin{equation} \label{eqn:urca}
    \begin{split}
        \isotm{Ne}{23}           &\rightarrow \: \isotm{Na}{23} + {e}^{-} + {\Bar{\nu}}_{e} \\
        \isotm{Na}{23} + {e}^{-} &\rightarrow \: \isotm{Ne}{23} + {\nu}_{e}
    \end{split}
\end{equation}
% ... 
The convective Urca process is the cyclical process that links the Urca
reactions with the convective mixing. The Urca reactions are mainly
density-dependent, and thus, within a WD, there are two distinct regions defined by one reaction dominating over the other. These regions are separated by the Urca shell, the spherical shell at which the Urca reactions are in local equilibrium. 
Convection continuously mixes material across the Urca shell, from dense interior regions where electron capture dominates, to less dense exterior regions where $\beta$-decay dominates, and vice versa.  
For the A=23 pair, the Urca shell is approximately located at $\rho_\text{urca} \sim 1.7 \times 10^9 $ $\gcc$ \citep{suzuki2016}, 

The cyclical nature of this process ensures that even a small 
fraction of Urca pair material can noticeably impact the WD’s 
evolution, and the convection zone itself \citep{lesaffre2005,stein-wheeler2006,piersanti2022}. 
This is done through the leakage of energy from the WD via neutrino emission and the formation of compositional gradients which stabilize regions to convection.
Additionally, the convective Urca process can impact the convection zone by producing small changes in the electron density, which can alter the density or pressure in the WD degenerate material. By affecting the density, the convective Urca process can produce a dragging effect on the convective mixing, resulting in slower mixing and/or a smaller convection zone, see Section \ref{subsec:buoy}.

This problem was previously addressed using an analytical-approach and 2D hydrodynamic simulations \citep{lesaffre2005, stein-wheeler2006}, showing that the convective Urca process may slow down convection. But, as more recent studies have shown (B2025), 3D hydrodynamic simulations are necessary in order to properly capture the effects of turbulent convection linked with nuclear reactions. Also, note that in this work we only consider the $A = 23$ Urca pair, but it has been suggested that the inclusion of other Urca pairs, like $\isotm{Ne}{21}-\isotm{F}{21}$ and $\isotm{Mg}{25}-\isotm{Na}{25}$, may have a significant impact on the results \citep{piersanti2022}.

\section{Low Mach Hydrodynamics: \maestro\ }
\label{sec:3}
The convection that arises in the simmering phase of a WD is slow compared to the sound speed (Mach number ${\sim} 10^{-3}$). To efficiently model the slow moving regime, we need to capture the fluid dynamics without having to resolve acoustic wave propagation. We use the \maestro\ low-Mach hydrodynamic code \citep{fan2019}, which is specially designed to model stellar interiors and atmospheres. \maestro\ is a massively parallel low Mach number hydrodynamic code built on the AMReX framework \citep{zhang2019}, that effectively filters out the sound waves while still accurately modeling the convection. In our simulations, a ``full-star" geometry is used, which places the star in the center of a 3D cartesian grid with an effective resolution of 5 km. 

The formulation of the equations
solved by \maestro\ relies on the existence of a base state density, $\rho_0$, and pressure, $p_0$, that are in hydrostatic equilibrium (HSE). The base state serves as a reference configuration around which perturbations evolve. In the simulations presented, unlike that in B2025, the base state is evolved. Meaning that \maestro\ solves the full equation set \citep{fan2019}, including relevant time derivatives of the base state:
\begin{gather}
    \label{eq:maestro1}
    \frac{\dif \rho X_k}{\dif t} + \nabla \cdot (\rho \rm{U} X_k) = \rho \dot{\omega}_k \\
    \label{eq:maestro2}
    \frac{\dif (\rho h)}{\dif t} = - \nabla (\rho h \rm{U}) + \frac{D\rho_0}{Dt} + \rho H_{nuc} \\
    \label{eq:maestro3}
    \frac{\dif \rm{U}}{\dif t} + \rm{U} \cdot \nabla \rm{U} + \frac{\beta_0}{\rho} \nabla \bigg(\frac{p '}{\beta_0}\bigg) = -\frac{\rho'}{\rho}|g|\rm{e}_r \\
    \label{eq:maestro4}
    \nabla \cdot (\beta_0 \rm{U}) = \beta_0 \bigg(S - \frac{1}{\Gamma_1p_0}\frac{\dif p_0}{\dif t}\bigg) \\
    \label{eq:maestro5}
    \nabla p_0 = -\rho_0 |g|\rm{e}_r
\end{gather}
where $\rho, p, \rm{U}, h$ are mass density, pressure, velocity and enthalpy respectively. $X_k$ is the mass fraction of the k-th isotope, $X_k \coloneqq \rho_k /\rho$, and $\dot{\omega}_k$ is the creation/destruction rate of that k-th isotope. Note that $X_k$'s are defined such that $\sum_{k}X_k = 1$. The base state is defined by $\rho_0$, and $p_0$, thus Eq.\ \ref{eq:maestro5} sets HSE. $\rho'$ and $p'$ are defined such that $\rho' \coloneqq \rho - \rho_0$ and  $p' \coloneqq p -p_0$. $H_{\rm{nuc}}$ accounts for the nuclear energy generation rate. $\beta_0$ is a density-like variable in the velocity constraint equation that captures the background stratification. Lastly, $S$ is the source term to the divergence constraint (Eq.\ \ref{eq:maestro4}) that accounts for perturbations related to compositional changes and heating from reactions, and $\Gamma_1$ is the lateral average of the first adiabatic exponent ($d \log p/ d \log \rho |_s$).
Note, for full-star geometries, as in the simulation presented here, the enthalpy equation is decoupled from the rest of the solution.
Instead, we define the temperature using the base state pressure, $p_0$, which ensures the basestate is in HSE and thermodynamic equilibrium.
The energy generated by nuclear reactions is coupled to the solution via the source term, $S$, in the velocity constraint (Eq.\ \ref{eq:maestro4}).

In our simulations, we solve a nuclear reaction network generated using the Python library \pynucastro\ \citep{smith2023}. The network incorporated accounts for simple carbon burning and the $A = 23$ Urca reactions. The reaction network is coupled to the fluid equations via Strang-splitting. We use a publicly available equation of state that is purposely built for degenerate stellar scenarios \citep{timmes2000, fryxell2000} \footnote{Further documentation and sources for the EOS can be found at \url{https://cococubed.com/code_pages/eos.shtml}}. It accounts for the contributions of photons, nuclei, electrons and positrons. The network and equation of state is implemented via the \microphysics\ project \citep{microphysics2024} \footnote{The full source code for \microphysics\ can be found at \url{https://github.com/AMReX-Astro/Microphysics}}. For further details on \maestro\ equations and algorithm see \cite{fan2019} or the full documentation \footnote{The \maestro\ documentation can be found at \url{https://amrex-astro.github.io/MAESTROeX/docs/index.html}}. Also, the full source code for \maestro\ can be found at \url{github.com/AMReX-Astro/MAESTROeX}, all contributions are welcome.

\section{Simulations}
\label{sec:4}
In order to characterize the effects of the convective Urca process, we ran full 3D \maestro\ simulations with and without considering the convective Urca process.
We disrupted the convective Urca process by leaving out the $\beta$-decay reaction. This way the carbon burning, which includes the electron capture reaction, is consistent between the simulations.
We denote the simulations as Full Network (FN1, FN2) and No Beta (NB1) with the numeral indicating the initial model. The initial models are presented in Subsection \ref{sec:4.1}.
In Subsection \ref{sec:4.2}, we describe the reaction networks used in these simulations.

\subsection{Initial Models} \label{sec:4.1}

We built two parameterized initial models using the framework developed and maintained by \citet{initial_models2024}. In building these models, we considered the results obtained in \citet{willcox2018} and B2025. 
Each model considered a 40\% - 60\% Carbon-Oxygen WD with a trace amount of Urca nuclei, 
$\Xisotm{Ne}{23} + \Xisotm{Na}{23} \approx 10^{-4}$, which is consistent with the start of the simmering phase as shown by stellar evolution models \citep{martinez-rodriguez2016, piersanti2017, schwab2017a}.
The total mass of the WD is 1.398 \Msun, meaning that it is a near Chandrasekhar-mass limit WD.
To construct the initial model we first set the central conditions, $T_c = 5.5 \times 10^{8}$ K and $\rho_c  = 4.5 \times 10^{9} \gcc$ as in B2025 and motivated by \cite{martinez-rodriguez2016}. 
We then define an isentropic core obtained by integrating outward from the central conditions while maintaining HSE and constant entropy.
We then transition from an isentropic profile to an isothermal profile, which should be stable to convection.

\begin{figure*}[ht!]
\plotone{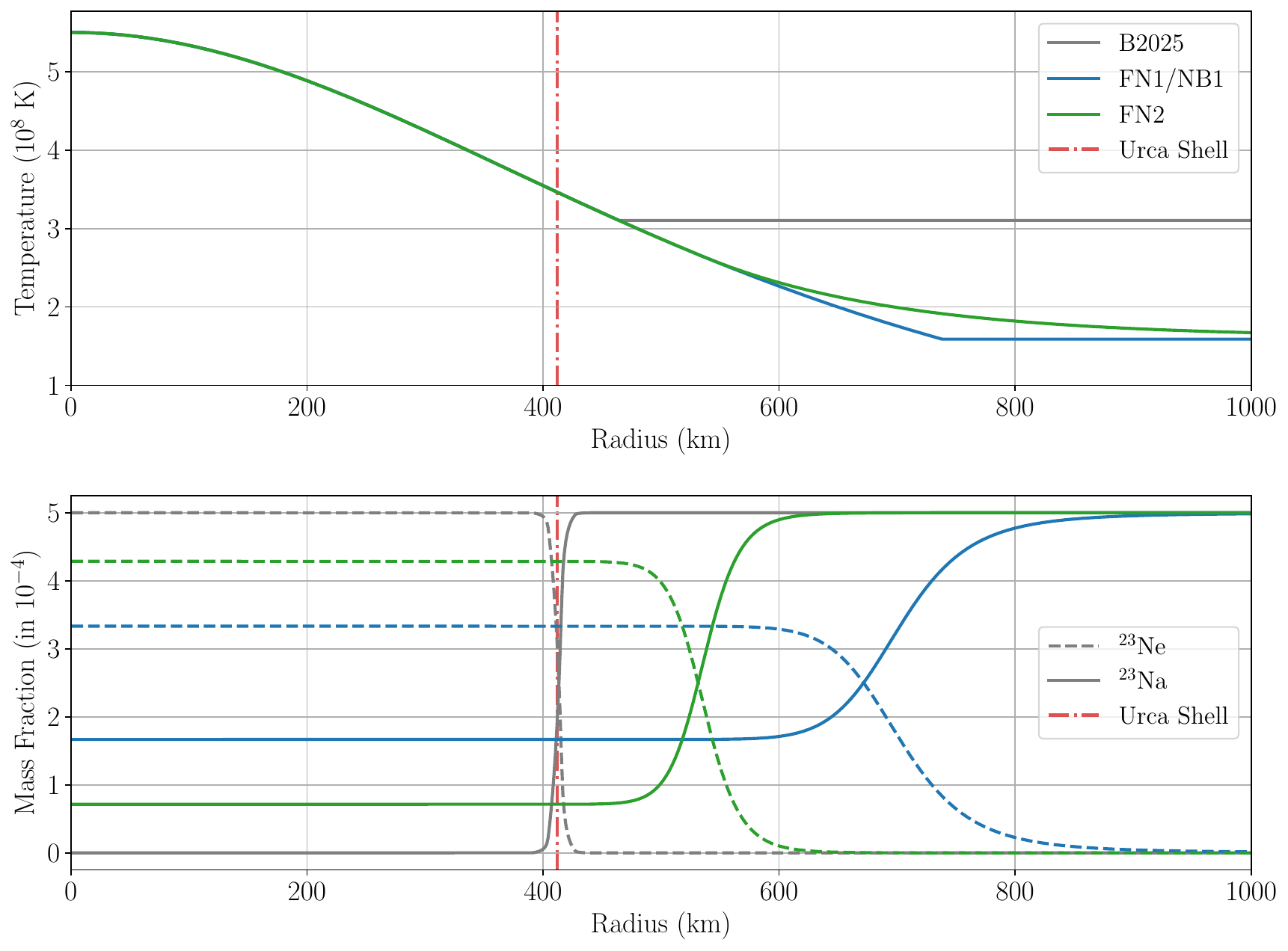}
\caption{Top plot: Star temperature vs radius profiles. Bottom plot: Mass fraction of
the Urca species vs radius. The grey curves represent the initial state of the simulations in B2025, the blue curves the initial state of the FN1 and NB1 
simulations, and the green curves the initial state of the FN2 simulation. The dashed curves represent the $\isotm{Ne}{23}$ and the solid ones the $\isotm{Na}{23}$. The vertical red line indicates the position of the Urca shell.}
\label{fig:init_models}
\end{figure*}

After careful consideration of the conclusions reached in B2025, we constructed a model with an isentropic core out to mass coordinate $1 \, \Msun$ or radius of 740 km (see top plot in Fig.\ \ref{fig:init_models}).
An isentropic core of this size aligns closely to the results of the fiducial model presented in \citet{martinez-rodriguez2016}.
The corresponding temperature of the isothermal region is $T_{\mathrm{out}} \approx 1.59 \times 10^8 \, \mathrm{K}$. 
The Urca pair are initialized closer to dynamic equilibrium, near the convective boundary, instead of around the Urca shell, as done in B2025 (see bottom plot in Fig.\ \ref{fig:init_models}). 
The initial models approximate the equilibrium state of the Urca pair assuming instantaneous mixing and the creation of \isot{Na}{23} due to carbon burning \citep{boydproceedings}. 
Starting with a composition near equilibrium helps the simulation efficiently relax to a steady state, saving computational time.
This initial model was used for both of the FN1 and NB1 simulations.

After running the FN1 simulation, we found that the well-mixed region was significantly smaller than the $1 \, \Msun$ isentropic region defined in the first initial model.
Thus, we constructed a new initial model which set the size of the isentropic core to $0.65 \, \Msun$ or radius of 538 km, in alignment with the results of FN1 (see Section \ref{subsec:mixing}).
In an attempt to keep a consistent outside temperature, we slowly decreased the temperature gradient such that the temperature trended to  $T_{\mathrm{out}} \approx 1.59 \times 10^8 \, \mathrm{K}$ as in the first initial model.
This slow decrease was achieved by multiplying the temperature gradient by a ``smoothing factor" ($f =0.9925$) at every discretized step beyond the end of the isentropic core (see green curve in top plot of Fig.\ \ref{fig:init_models}). 
Note that although we do not have a strictly isothermal envelope in this initial model, the temperature profile outside of 538 km should still be stable to convection.
For this second model, we distributed the Urca pair to align with the new, smaller isentropic core (see bottom plot of Fig.\ \ref{fig:init_models})
We discuss the drawbacks and likely unphysical nature of this model in Section \ref{sec:6}, and our goals in future work to start from more realistic temperature profiles.

\subsection{Nuclear Network} \label{sec:4.2}

For this study, we built a nuclear network using \pynucastro. We considered a
simple carbon burning chain, the neutron decay, the proton plus electron capture, and the $A = 23$ Urca reactions. The rates
for the carbon burning reactions were obtained from the JINA REACLIB database \citep{cyburt2010} and the
tabulated rates for the Urca reactions from \citet{suzuki2016}. This nuclear network is slightly different from the one considered in B2025. We have now incorporated the $p + e^{-} \rightarrow n$ reaction, and we use the rates presented in \citet{langanke_pinedo2001} for the $n \leftrightarrow p$ reactions, which account for density dependence. Along with Urca reactions in Eq.\  \ref{eqn:urca}, the network includes the following rates:
\begin{equation}
     \begin{split}
        \isotm{C}{12}  + \isotm{C}{12} &\rightarrow \: \isotm{Ne}{20} + \isotm{He}{4}   \\
        \isotm{C}{12}  + \isotm{C}{12} &\rightarrow \: \isotm{Mg}{23} + n               \\
        \isotm{C}{12}  + \isotm{C}{12} &\rightarrow \: \isotm{Na}{23} + p               \\
        \isotm{C}{12}  + \isotm{He}{4} &\rightarrow \: \isotm{O}{16}                    \\
        n                              &\rightarrow \: p + {e}^{-} + \bar{\nu}_e \\
        p + e^{-}                             &\rightarrow \: n + \nu_e
    \end{split}   
\end{equation}

To remove the cyclical effects of the convective Urca process in the NB1 simulation, we only need to remove one of the two reactions involving the Urca pair. 
Since the electron capture reaction is directly linked with the carbon burning process because the $\isotm{Na}{23}$ electron captures just after being produced in the $\isotm{C}{12}$ burning chain, we remove the $\beta$-decay reaction: $\isotm{Ne}{23}  \rightarrow \: \isotm{Na}{23} + {e}^{-} + {\nu}_{e} $.

\section{Results} \label{sec:5}
We ran all three simulations for about an hour or longer of simulation time, which equates to tens of convective turnovers. 
The FN1 simulation reached a quasi-steady state with a split convection zone or large, potentially unphysical, overshooting region.
The FN2 and NB1 simulations relaxed to a steady state based on the size of the convection zone and the stabilization of the rms velocity in said convection zone. 
We compare the convective structure and kinematics of all three simulations in Subsection \ref{subsec:mixing}, then focus our comparisons of the energy output between the two steady-state simulations, FN2 and NB1, in Subsection \ref{subsec:energy}.
In Subsection \ref{subsec:buoy}, we analyze the Urca reactions in relation to the buoyancy which may disrupt the convective motions.
And in Subsection \ref{subsec:burn}, we display the evolution of the carbon burning for each simulation.

\subsection{Convective Structure and Kinematics}
\label{subsec:mixing}
\begin{figure*}[ht!]
\plotone{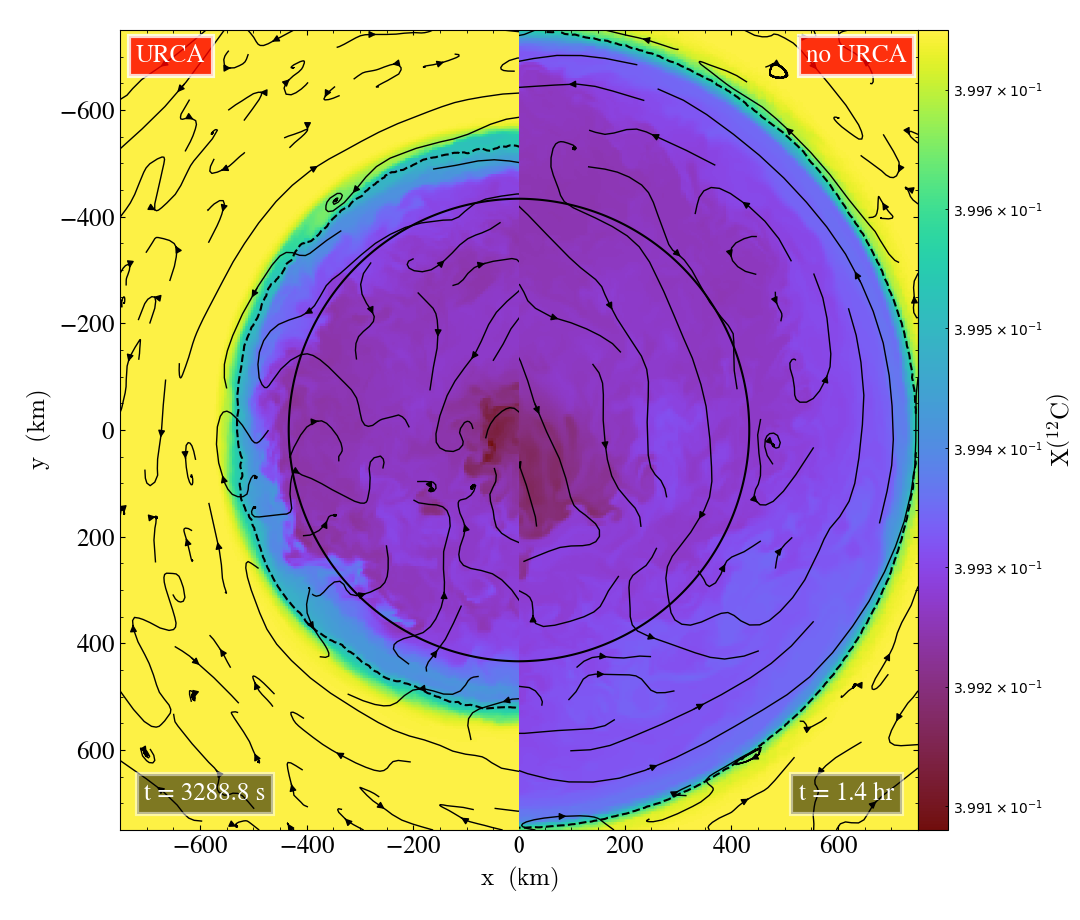}
\caption{
 2D slice through the center of the WD colored by the value of 
 $\Xisotm{C}{12}$ at the
end of the simulation. The solid and dashed black lines represent the Urca shell and
the boundary where $\Xisotm{Na}{23} = \Xisotm{Ne}{23}$ respectively. The black streamlines indicate the trajectories of the convective flows. The left slice represents the
FN2 simulation, and the right one the NB1 simulation.
}
\label{fig:slices}
\end{figure*}

\begin{figure}
    \centering
    \plotone{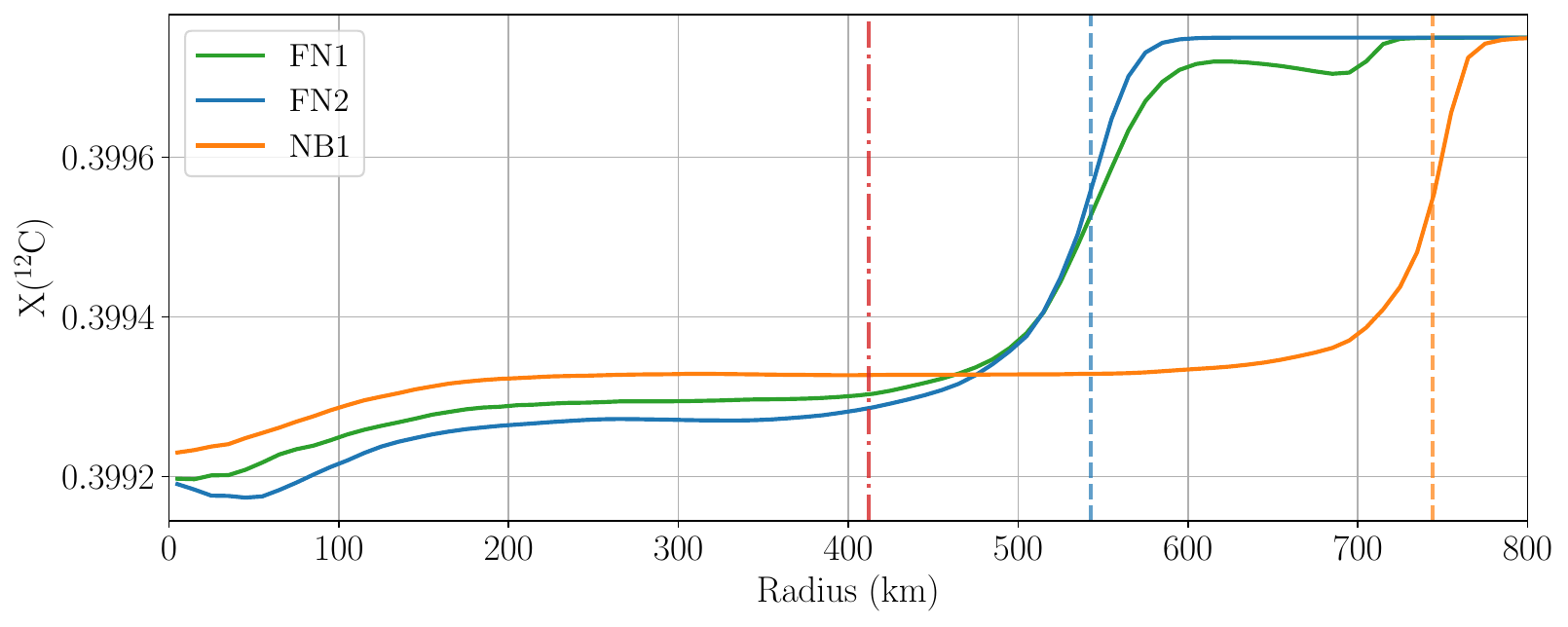}
    \caption{Spherically averaged \Xisot{C}{12} profiles vs radial bin for each simulation. The green curve is from FN1 simulation. The blue curve from the FN2 simulation. And the orange curve from NB1 simulation. The red vertical dash-dot line indicates the location of the Urca Shell. The vertical dashed lines indicate the \Rconv\ for the FN2 (blue) and NB1 (orange) simulations}
    \label{fig:all_sims_c12prof}
\end{figure}

For each simulation, we initialize a relatively small velocity field, maximum magnitude of $1 \, \kms$, as described in B2025 and more extensively in \citet{zingale2009}. 
These small velocity perturbations are mainly helpful in the initial projection of the velocity field to the divergence constraint (see Eq.\ \ref{eq:maestro4}), and are quickly replaced by the convective motions.
The carbon burning near the center of the star begins to heat material and drive convective plumes outward, mixing the ash in the outer regions. 
This establishes a convection zone which we define by the extent to which carbon burning ash is mixed. 
We follow a similar definition as B2025, where \Rconv\ is defined as the radius at which the \Xisot{C}{12} profile's radial gradient is maximal.
Motivation for this definition can be seen in Fig.\ \ref{fig:slices} where we have plotted the slices of the FN2 and NB1 \Xisot{C}{12} at the end of each simulation.
Additionally, we plot the 1D profiles of \Xisot{C}{12} in Fig.\ \ref{fig:all_sims_c12prof} for each simulation.
With this defined, we can then track the mass enclosed by the convection zone, \Mconv, and the rms velocity, $\vrmsm =\sqrt{<U_x^2> + <U_y^2> + <U_z^2>}$, where the $<>$ represent averages over all cells in the convection zone (the same definition as in B2025).

\begin{figure}[ht!]
    \centering
    \plotone{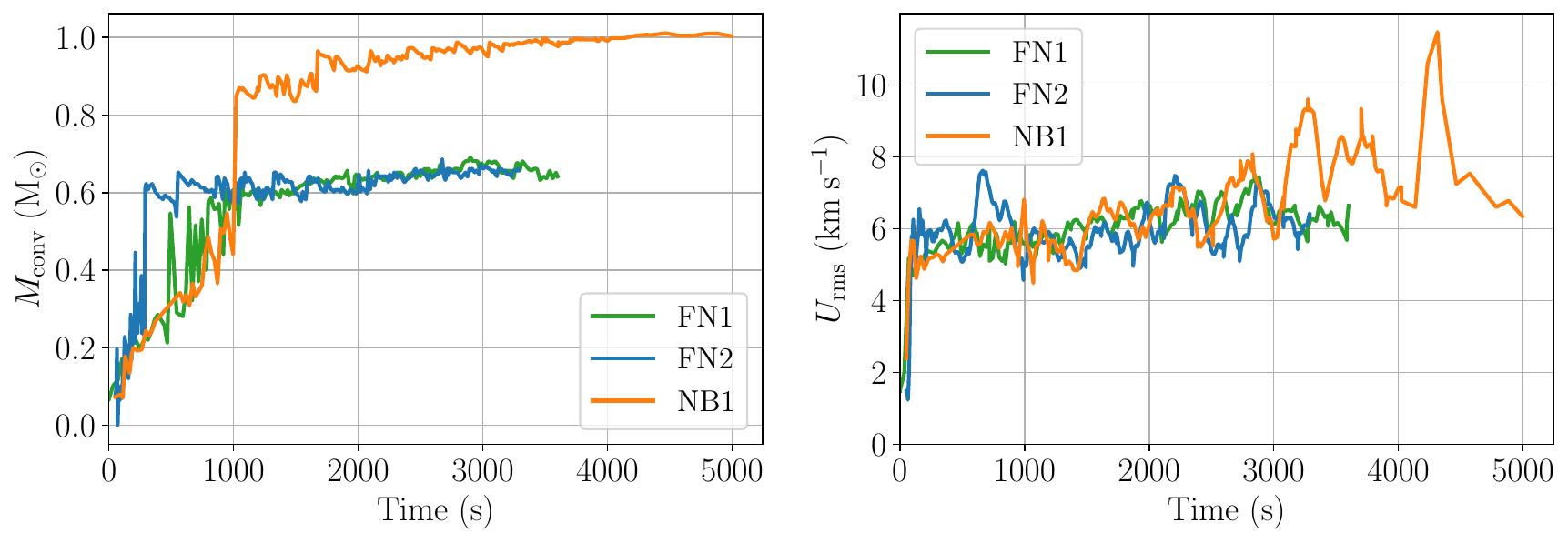}
    \caption{The left plot tracks the mass contained in the convection zone, \Mconv, over simulation time. The right plot shows the \vrms\ of the convection zone over time. We plot all three simulations with FN1 in green, FN2 in blue and NB1 in orange.}
    \label{fig:mconv_vrms}
\end{figure}

In Fig.\ \ref{fig:mconv_vrms}, we see a convection zone is largely established within the first ${\sim}1000 \, \mathrm{s}$ for each simulation, even sooner for FN2. 
After this initial transient period, FN1 and FN2 relax to a stable convective mass and the convective velocity remains around ${\sim}6 \, \kms$. 
The NB1 simulation takes a longer time to relax to a steady convective mass, which explains why the simulation was run 50\% longer than the FN1/FN2 simulations.
The NB1 convection zone is nearly twice the size of the FN1/FN2 simulations; see Fig.\ \ref{fig:slices} for further demonstration.
In terms of convective velocities, the NB1 \vrms\ is slightly higher though also oscillates more than the FN1/FN2 simulations, whose \vrms\ are roughly equivalent.

The FN1 simulation differs substantially in structure compared to FN2 and NB1 simulations. 
This is easily seen in comparing \Xisot{C}{12} profiles, see Fig.\ \ref{fig:all_sims_c12prof}.
The FN2 and NB1 simulations, blue and orange curves, have a clear convective core with a convective boundary which smoothly, and monotonically, transitions to a stable envelope.
The FN1 simulation, green curve, has a convection zone that extends to about 545 km, and then an additional partially mixed zone from roughly 545 to 720 km.
In Section \ref{sec:6}, we discuss the physical meaning of this region and how it may relate to a realistic overshooting region on the convective boundary.

In the FN1/FN2 simulations, the Urca nuclei relax to a similar equilibrium distribution, despite starting from very disparate initial models (bottom plot of Fig.\ \ref{fig:init_models}). 
This equilibrium distribution is based primarily on the size of the electron capture region and $\beta$-decay regions, with an additional source term for \isot{Na}{23} from the carbon burning. 
Since there is no $\beta$-decays in the NB1 simulation, the equilibrium distribution of the Urca pair is based solely on the generation of \isot{Na}{23} from carbon burning balancing the electron capture rate to \isot{Ne}{23}.

\begin{figure}[ht]
    \centering
    \plotone{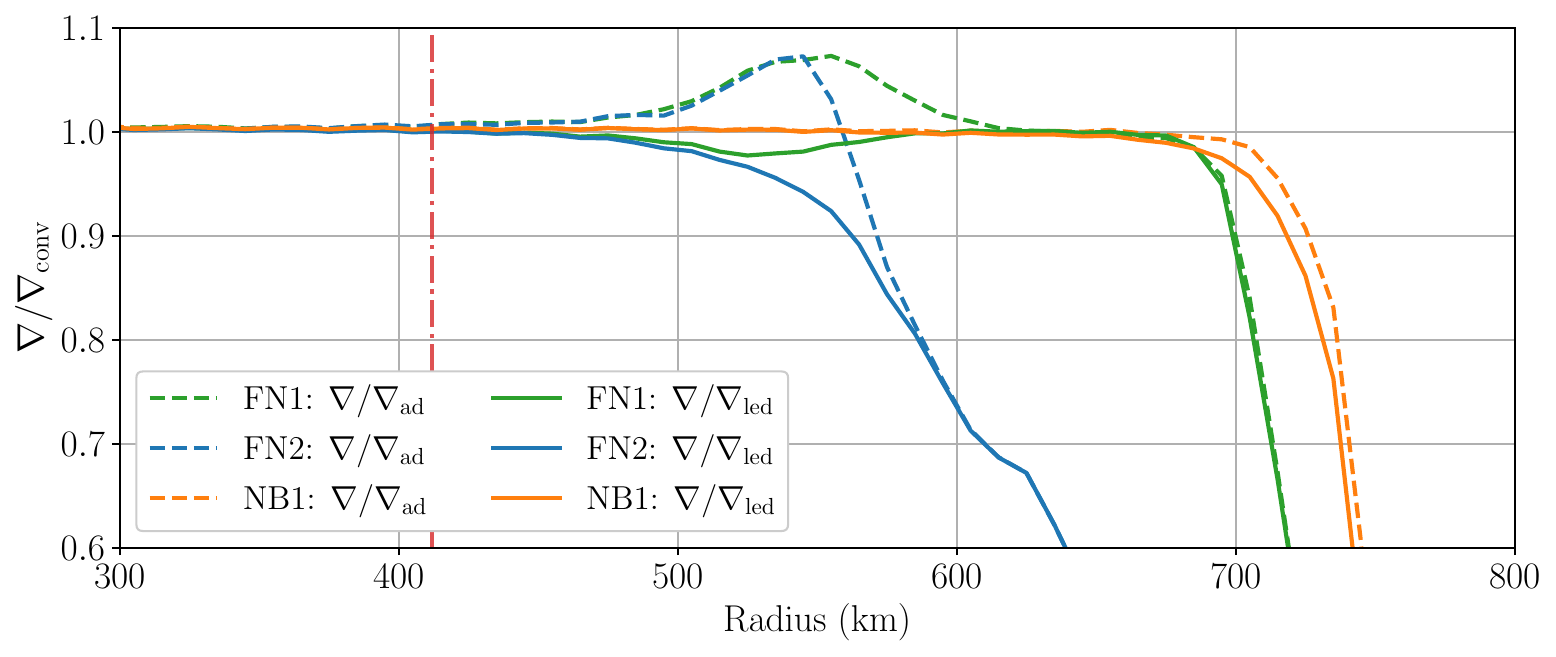}
    \caption{Spherically averaged convective gradient profiles vs radial bin for each simulation. The dashed curves represent the ratio of the real gradient to an adiabatic $(\nabla / \nabla_{\mathrm{ad}}$. The solid curves represent the ratio of the real gradient to the Ledoux gradient $(\nabla / \nabla_{\mathrm{Led}}$. The green curve is from FN1 simulation. The blue curve from the FN2 simulation. And the orange curve from NB1 simulation. The red vertical dash-dot line indicates the location of the Urca Shell.}
    \label{fig:all_sims_conv_criteria}
\end{figure}

To further investigate the convective boundaries of these simulations, we consider their temperature gradients.
The temperature gradients of stellar interiors are considered stable to convection based on the Schwarzschild or Ledoux criterion \citep{HKT2004}.
The Ledoux criterion incorporates the stabilizing effects of compositional gradients, and reduces to the Schwarzschild criterion in the absence of any such compositional gradients.
In Fig.\ \ref{fig:all_sims_conv_criteria}, we plot the ratio of the temperature gradient, $\nabla = d \log T / d \log P$,  compared to the adiabatic gradient $\nabla_{\mathrm{ad}}$ (dashed curves) and the Ledoux gradient  $\nabla_{\mathrm{Led}}$ (solid curves).
We calculate the adiabatic and Ledoux gradients equivalent to B2025, following the prescription in \citet{paxton2013}.
Looking at the Schwarzschild criterion, regions where $\nabla/\nabla_{ad} > 1$, or very near 1, should be convectively unstable, while those less than one should be stable to convection. 
In the FN1 and FN2 simulations, we see a super-adiabatic region peaking around 550 km.
This aligns well with the $R_\text{conv}$ of each of these simulations. 
And if we look at the Ledoux criterion, we find that the compositional gradients are stabilizing this region around 550 km.
Outside 550 km, the FN1 and FN2 simulations differ, in part due to their respective initial models.
The FN1 simulation returns to following an adiabatic gradient until about 675 km, where the gradient transition to the isothermal, convectively stable, region.
In contrast, the FN2 simulation remains convectively stable from $550 \, \mathrm{km}$ outward as it smoothly transitions to the sub-adiabatic gradient.
In each case, the temperature gradient outside ${\sim} 550 \, \mathrm{km}$ aligns with the initial models presented in Fig.\ \ref{fig:init_models}.

The NB1 simulation largely remains consistent with the set initial model of an isentropic region out to about 750 km. 
Looking at the Ledoux criterion, there is a slightly stable region beginning at about 650 km, due to the compositional gradient at the convective boundary.
However, due to the lack of the convective Urca process and any $\beta$-decays in NB1, there is significantly less compositional stability being provided.
Thus there is only relatively small differences between $\nabla/\nabla_{\mathrm{ad}}$ and $\nabla/\nabla_{\mathrm{Led}}$ for this simulation.

Note, the region of clear convective stability in FN1, ${\sim}540 \, \mathrm{km}$, is well outside the Urca shell which is located at about 412 km.
The Urca reactions near the Urca shell are very slow compared to the convective mixing near the Urca shell. 
However, further out from the shell, at lower densities, the $\beta$-decay reactions are much faster and can much more significantly affect convection.

\begin{figure}[ht!]
\plotone{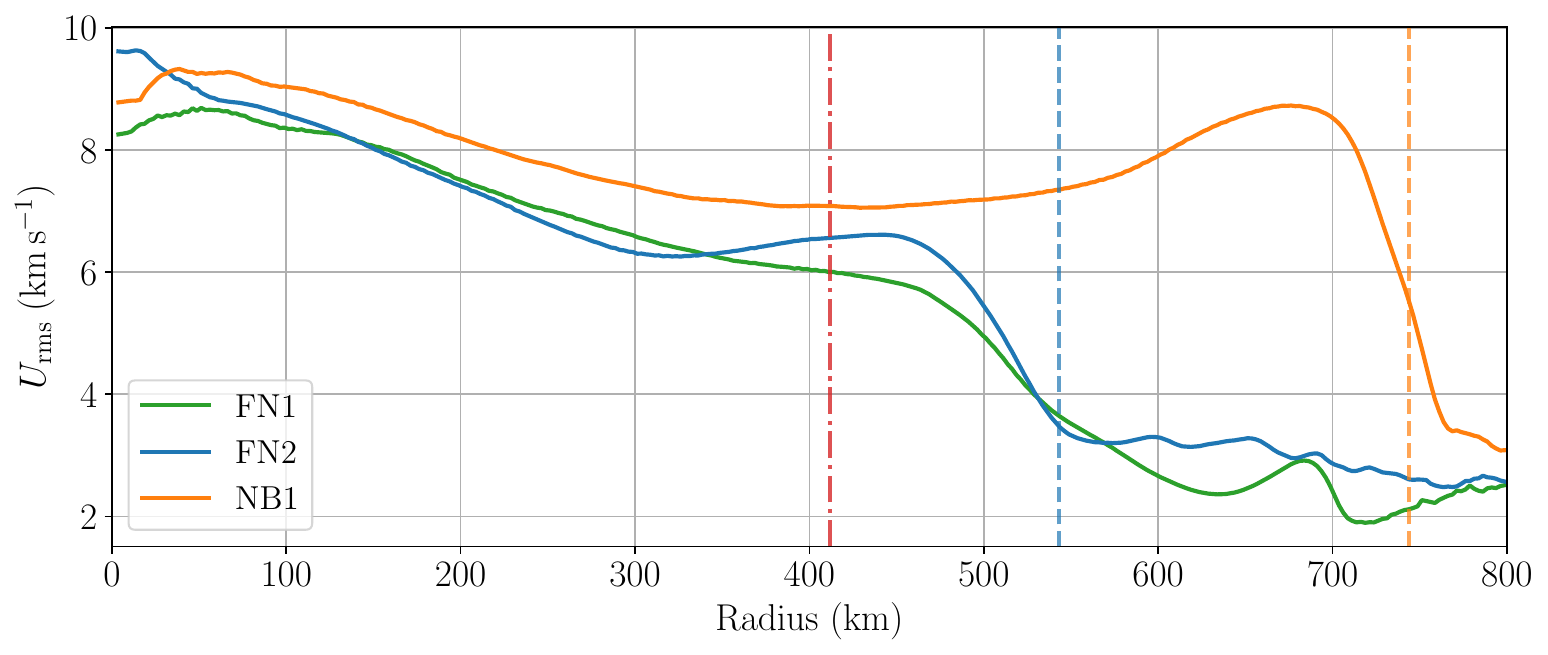}

\caption{ \vrmsr\ against radius profiles averaged over the time. That is the last $400$ s for the FN1 simulation (represented by the green curve), $400$ s for the FN2 simulation (represented by the blue curve), and the last $500$ s for the NB1 simulation (represented by the orange curve). The vertical dashed lines indicate the \Rconv\ for the FN2 (blue) and NB1 (orange) simulations. The red vertical dash-dot line indicates the location of the Urca Shell.}
\label{fig:urms}
\end{figure}

Looking at the velocity structure of the convection zone is also informative. 
However, the velocities are inherently turbulent and prone to large fluctuations.
Thus we average over a few convective turnover times.
We calculate the convective turnover timescale as $\tau_\text{conv} = 2R_\text{conv}/U_\text{rms}$. 
We obtain similar timescales for each simulation, for FN1/FN2 $\tau_\text{conv} \approx 190 $ s and for NB1 $\tau_\text{conv} \approx 215 $ s. 
We average over 400 s for FN1/FN2 and 500 s NB1 when plotting \vrmsr\ profiles (Fig.\ \ref{fig:urms}).
Note that these \vrmsr\ profiles are calculated similar to that in Fig.\ \ref{fig:mconv_vrms}, except the averages are over radial bins centered at radius $r$ instead of the full convection zone.
As already discussed and presented in Fig.\ \ref{fig:mconv_vrms}, the NB1 simulation has a slightly higher \vrms.
This is also reflected in the \vrms\ profiles where though there is a similarly shaped distribution, the NB1 has higher velocities, along with extending out to ${\sim} 745 \, \mathrm{km}$.
With the larger convective velocity, and the much larger convection zone, the NB1 has around 40\% higher kinetic energy than FN2, see Table \ref{tab:sum}.

\begin{figure}[ht]
    \centering
    \plotone{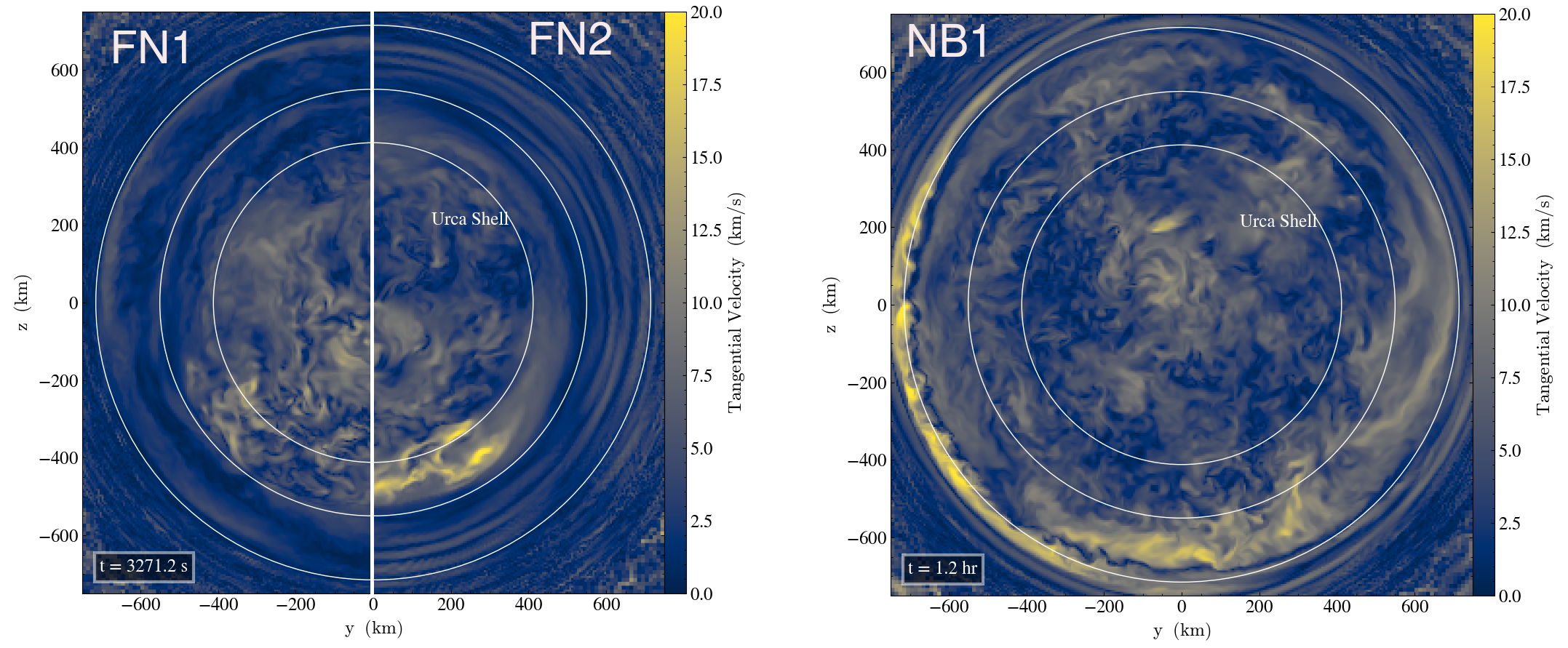}
    \caption{Slices through the center of the white dwarf, zoomed into the convective core region. From left to right we have the FN1, FN2, and NB1 simulations. The inner white circle indicates the Urca shell. The outer two circles approximately mark the edge of the convection zones for the two simulations 545 km and 745 km respectively.}
    \label{fig:tanvel}
\end{figure}

Along with the \vrmsr\ profiles, it is informative to look at a 2D slice of the velocity distribution.
In Fig.\ \ref{fig:tanvel}, we plot the tangential velocity, $U_\text{tan} = | \textbf{U} - U_r |$ where $U_r$ is the radial velocity component.
Comparing FN1 and FN2 directly, the left plot, we see that although the bulk of the velocity is contained to $R< 540$ km, the FN2 has a clearly stable region which displays wave-like motions. 
For the FN1 simulation, this region from 540 to 700 km displays noticeably slow velocities but still relatively chaotic/turbulent flow. 
In contrast, the NB1 simulation clearly shows a fully filled out convection zone out to 745 km.

Note, in all three simulations, the outer regions of $R> 850$ km (see the corners of Fig.\ \ref{fig:tanvel}) have what appears to be near random velocities. 
We believe these are due to small-scale excitations that we do not resolve. 
These excitations occur only in the convectively stable outer regions and do not effect the interior convective motions. These issues also worsen with decreased resolution, which again indicates we are under-resolving these wave oscillations.

In summary, both FN2 and NB1 simulations reached a steady-state defined by a convective core of a size roughly similar to the initialized isentropic region, \Rconv\ of 543 km and 744 km respectively. 
That translates to convective cores with masses of about $0.66 M_\odot $ and $1.0 M_\odot$.
The FN1 simulation had a different structure with a convection zone much smaller than the initial isentropic region, which resulted in a large partially mixed envelope.
Due to this behavior, we determine FN1 is only in a quasi-steady state as the evolution of the semi-mixed region is unclear.
The location of the FN1 convective boundary was used as motivation to construct the initial model used in FN2, see Section \ref{sec:4.1}
Aligning with the results shown in B2025, we see that, even though the convective Urca process results in a smaller convective core for the conditions considered, convection is not strictly constrained to the Urca shell as was suggested in \citet{stein-wheeler2006}.

\subsection{Energy Generation and Neutrino Losses} \label{subsec:energy}
We restrict this analysis to the FN2 and NB1 simulations because they reached steady state and had clear convective boundaries.
When using full-star geometries with \maestro, the enthalpy is decoupled from the rest of the solution and can slowly drift from thermodynamic equilibrium \citep{malone2011}, 
Thus, we focus our analysis on the instantaneous energy generation rate.
We track the energy generation rate due to reactions split into three components. That is the nuclear energy, \epsnuc, (change in relative binding energies), the neutrino emission, \epsnu, which free-streams from the WD, and the change in energy due to the compositional shifts, \epstherm. We dub this last term the thermal contribution and it is primarily important due to the capture/emission of highly degenerate electrons. The relatively complex thermal properties of the Urca reactions have been extensively studied in prior works \citep{bruenn1973, barkat_wheeler1990, piersanti2022} and we provide a brief summary in Appendix \ref{sec:fermi}.

In solving the nuclear network, we evolve energy at the same time as the change in isotopes.
We save this nuclear energy, which accounts for the change in relative binding energies between the isotopes and neutrino losses (from thermal emission and weak reactions).
To calculate \epsnuc, we subtract off the neutrino contributions.
\epsnu\ is calculated using the tabulated weak rates, similarly to when we solve the nuclear network. Thermal neutrinos are also included in this term, though they don't significantly contribute in our regions of interest.
For the thermal term, we calculate the energy due to compositional shifts using the creation/destruction rates from solving the nuclear network and the equation of state to calculate $\partial e / \partial X_k$. Here, $e$ is the internal energy, and the exact formula we use to compute $\partial e / \partial X_k$ is:
\begin{equation}
    \left.\frac{\partial e}{\partial X_k}\right|_{\rho, T, (X_j, j\neq k)} = \left.\frac{\partial e }{\partial \bar{A}}\right|_{\rho, T, \bar{Z}} \frac{\bar{A}}{A_k}  (A_k - \bar{A}) + \left.\frac{\partial e}{\partial \bar{Z}}\right|_{\rho, T, \bar{A}} \frac{\bar{A}}{A_k} (Z_k - \bar{Z})
\end{equation}
where $A_k$ and $Z_k$ are respectively the mass number and the atomic number of the k-th isotope, and $\bar{A}$ and $\bar{Z}$ are the averaged values in the cell.
Note, the primary contributions to this expression come from the Fermi energy associated with the electrons, as can be seen in Appendix \ref{sec:fermi}. 
The thermal component is then computed as:
\begin{equation}
   \dot{\epsilon}_{\mathrm{therm}}  = \sum_k \frac{\partial e}{\partial X_k} \dot{\omega}_k
\end{equation}

\begin{figure}[ht!]

\plotone{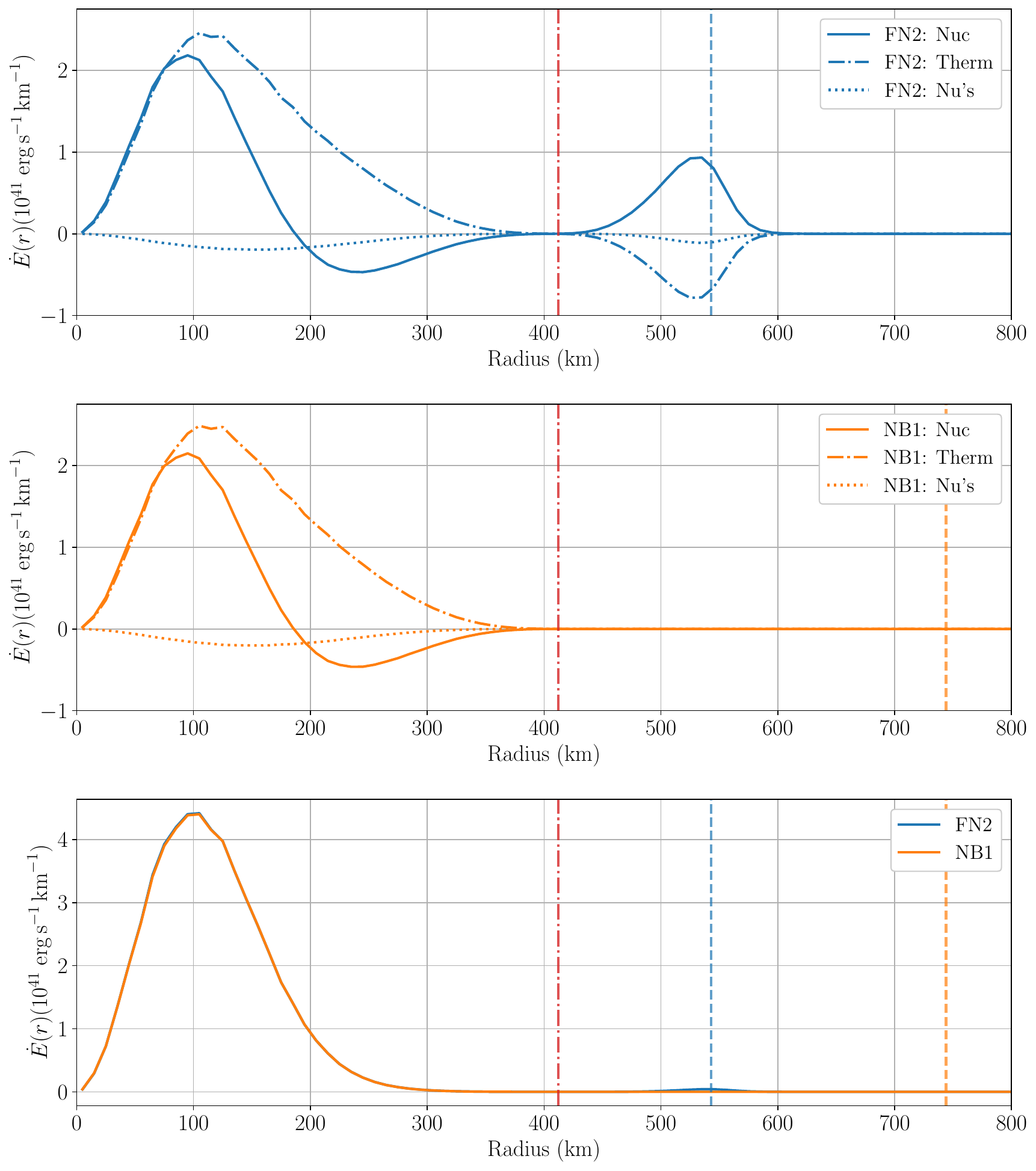}
\caption{
Top plot: Energy generation rate components per radial bin in the FN2 simulation. Middle plot: Energy generation rate components per radial bin in the NB1 simulation. Bottom plot: Total energy generation rate per radial bin with all the contributions summed up. The blue lines represent the FN1 simulation, the orange
lines the NB2, and the vertical lines indicate the convective boundaries.
}
\label{fig:energy}
\end{figure}

In the first two plots of Fig.\ \ref{fig:energy}, we plot the profiles of \epsnuc, \epsnu, and \epstherm. 
The top plot describes the energies from the FN2 simulation at the end of the simulation.
The middle plot does the same for the NB1 simulation, but at a time ($t=4350 \, \mathrm{s}$) prior to the end of the simulation.
We have chosen these times to achieve comparable carbon burning rates (see Subsection \ref{subsec:burn}). 
Note that in each case we are still in the steady state phase of the final few convective turnovers.
Keeping the carbon burning relatively constant allows us to more clearly isolate the effects of the Urca reactions.
This is a similar logic to why the electron capture rate is still included in the NB1 simulation, it is a key rate in the carbon burning process.

Concerning the energy profiles for the \epsnuc\ term, we distinguish a first peak around radii $\sim 100 $ km which represents the energy being generated by the carbon burning reactions. Then, at around radii $\sim 250$ km, we observe a dip in the profile, which corresponds to energy being consumed due to the electron-captures. Lastly, beyond the Urca shell, the profile shows a peak due to the energy generated by the $\beta$-decays. Note that we only see this last peak for the FN2 simulation since we removed this reaction in the NB1 simulation. 
The total nuclear energy generation rates for the simulations are $\dot{E}_{\rm nuc, \; FN2} = 2.38 \times 10^{43} \erg~\second^{-1}$ and $\dot{E}_{\rm nuc, \; NB1} = 1.67 \times 10^{43} \erg~\second^{-1}$. 

Considering the profiles of the energy lost due to neutrino emission, \epsnu, we observe two dips corresponding to the electron-capture reaction (central regions) and the $\beta$-decay reaction (outer regions). 
Again, the second dip is absent in the NB1 simulation. 
The total rates of energy losses due to $\nu_e$ emission are 
$\dot{E}_{\nu_{e}, \; {\rm FN2}} = {-}4.25 \times 10^{42} \erg~\second^{-1}$ and 
$\dot{E}_{\nu_{e}, \; {\rm NB1}} = {-}3.61 \times 10^{42} \erg~\second^{-1}$, showing the FN2 simulation loses ${\sim}20\%$ more energy due to neutrino emission than the NB1 simulation. 
The losses due to neutrino emission represent energies of about an order of magnitude less than the energy generated by carbon burning. 
This result also aligns with the results presented in B2025, where the total neutrino losses were estimated to represent only about $12\%$ of the energy generated by the purely nuclear term. 

Lastly, considering the \epstherm\ or compositional shifts contribution, we observe a peak in the central  of the star, and a dip in the region outside the Urca shell for the FN2 simulation. 
This mirrors the change in electron fraction due to reactions, as the electrons are the primary contribution to this term, see Appendix \ref{sec:fermi}.
As in the other energy contributions, there is no outer peak for the NB1 simulation.
In the interior, the thermal energy of the captured electron is left behind in the medium when this \epstherm\ term is considered separately, leading to a positive distribution of energy remaining in the medium.  
For the $\beta$-decay, energy must be provided to the electron emerging into the medium, thus it appears as energy lost from the overall medium.  
Each of these is offset to some degree by the corresponding nuclear process in each case (electron capture taking part or $\beta$-decay providing part) as discussed in Appendix \ref{sec:fermi}.
The total thermal energy rates for the simulations are $\dot{E}_{\rm therm, \; FN2} = 3.48 \times 10^{43} \erg~\second^{-1}$ and $\dot{E}_{\rm therm, \; NB1} = 4.09 \times 10^{43} \erg~\second^{-1}$. 

In the bottom plot in Fig.\ \ref{fig:energy}, we show profiles of the total energy generation rate accounting for all three energy components.
We observe that there are only slight differences between the energy profiles from both simulations.
This is primarily because the \epstherm\ and the \epsnuc\  terms largely cancel each other when it comes to the Urca reactions, which can be best seen in the outer region.
As a result, the energy generation is dominated by the central carbon burning.
The total energy generation rates are $\dot{E}_\text{FN2} = 5.44 \times 10^{43}\erg~\second^{-1}$ and $\dot{E}_\text{NB1} = 5.40 \times 10^{43}\erg~\second^{-1}$, meaning that there is only marginally more energy generated in the FN2 simulation with the full convective Urca process.
This aligns with previous results, where it was shown that the neutrino losses from the convective Urca process do not provide net cooling. 

\subsection{Buoyancy and Electron Fraction} \label{subsec:buoy}

As discussed in Section \ref{subsec:mixing}, more material is being moved and at a slightly higher averaged velocity in the NB1 simulation, resulting in a higher kinetic energy for this simulation compared to FN2. 
In contrast, we found in Section \ref{subsec:energy} that slightly more energy is generated in the FN2 simulation.
Knowing that the main difference between simulations is that we removed the cyclical Urca reactions, we can conclude this discrepancy is due to the effects of the convective Urca process. 
In particular, the weak reactions alter the buoyancy of the fluid and works against the convective mixing \citep{couch1974, iben1978, mochkovitch1996, sbw1999}.
The way the convective Urca process alters the convective mixing has primarily been implemented in 1D simulations via a work term that accounts for the work done moving electrons up the electron chemical gradient \citep{iben1978, iben1982, piersanti2022}.
In this section, we explore the comparable explanation \citep{couch1974} that the weak reactions affect the buoyancy, resulting in a drag on convection.

\begin{figure}[ht!]
\plotone{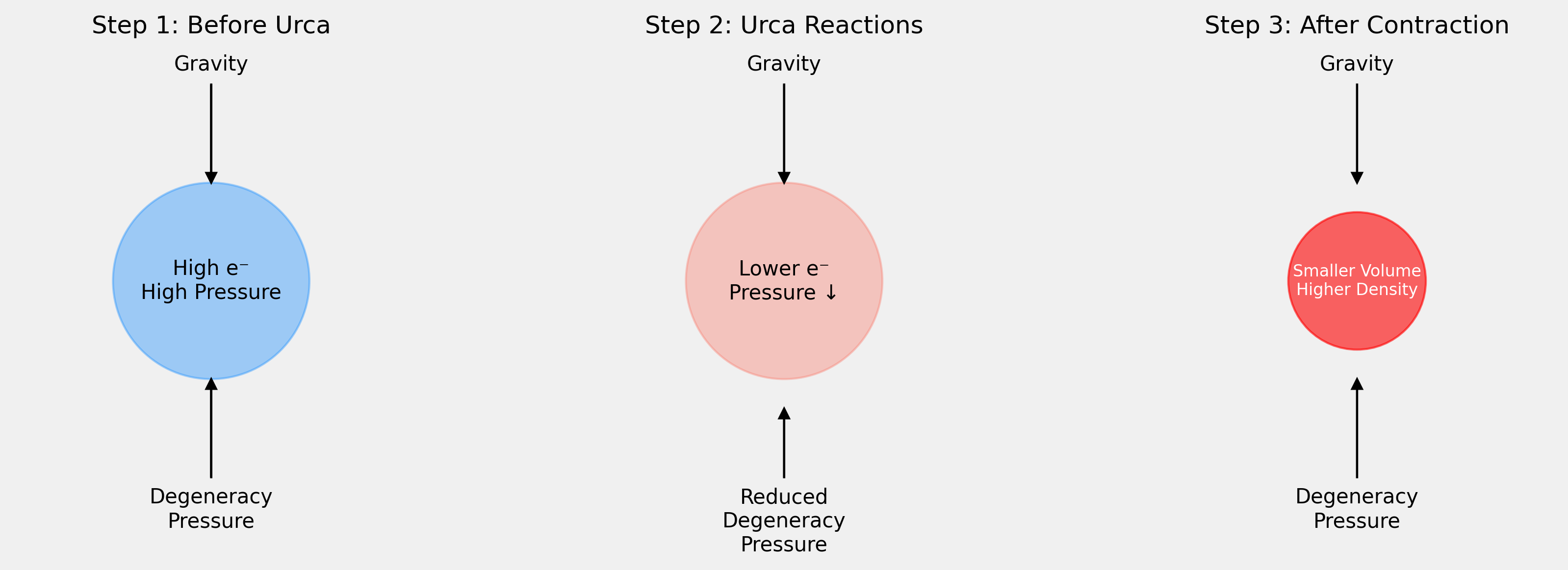}
\caption{
Schematic illustration of the simplified effects of the Urca reactions in a bubble of degenerate material.
}
\label{fig:scheme}
\end{figure}

A WD is an electron degenerate star, meaning that the gravitational forces are mainly balanced by the electron degeneracy pressure. 
The Urca reactions can directly alter the electron fraction in the convection zone, which leads to a necessary change in the density to maintain the degeneracy pressure in balance with gravity.
A schematic of this effect can be seen in Figure \ref{fig:scheme}.

\begin{figure}[ht!]
\plotone{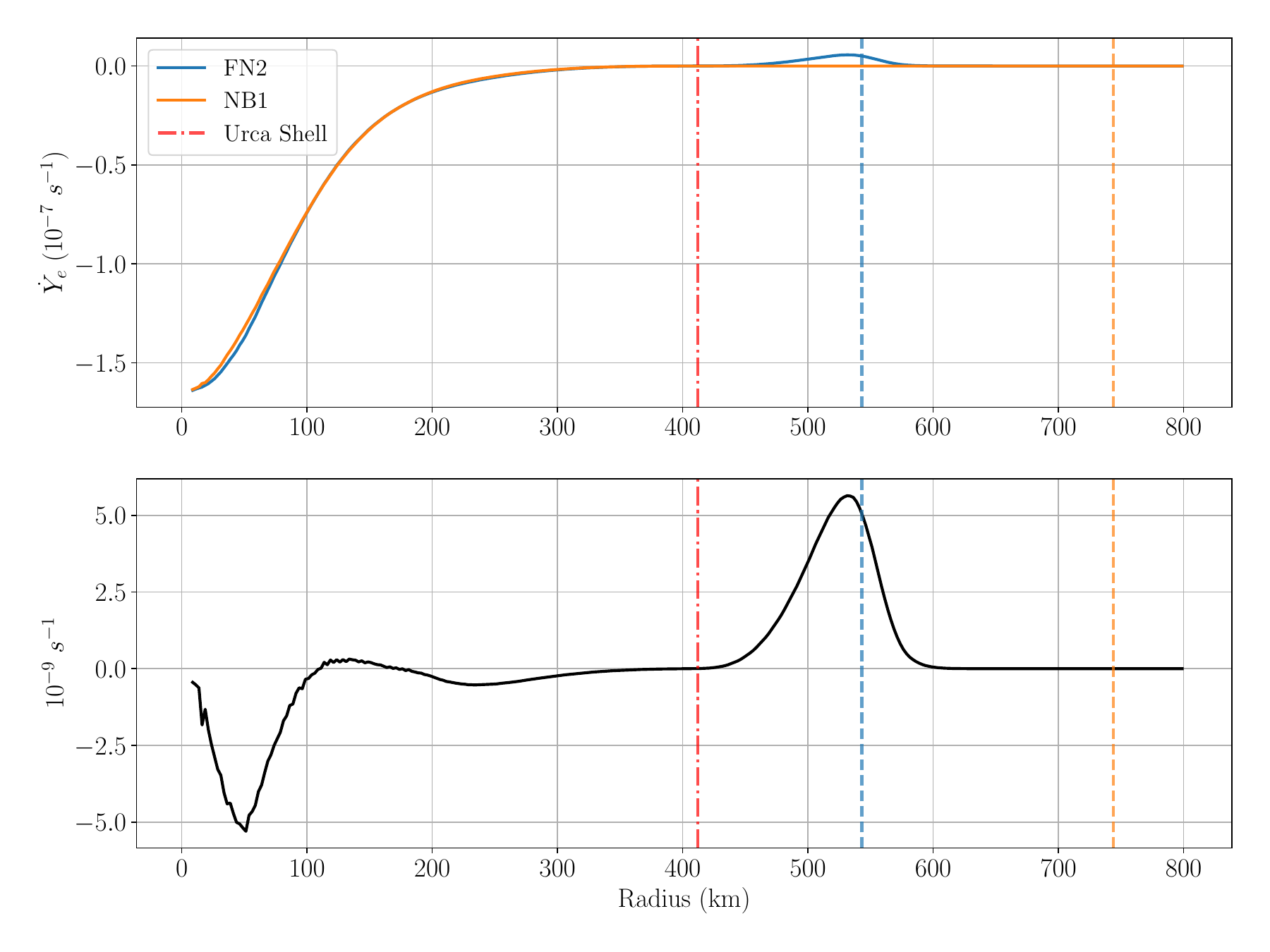}
\caption{
Top plot: Electron fraction rate. The blue curves represents the FN2 simulation that considers the Urca reactions, and the orange ones the NB1 simulation that excludes $\beta$-decays. Bottom plot: The difference between the electron fraction rate profiles plotted above. The vertical red line indicates the presence of the Urca shell, and other vertical dashed lines indicate the boundary of the convective region in each simulation.
}
\label{fig:electron}
\end{figure}

In Fig.\ \ref{fig:electron}, we plot the average change in electron fraction, $\dot{\mathrm{Y}}_e(r)$ vs radial bin. 
The main change in electron fraction is a product of the central \isot{C}{12} burning seen by the large dip in the inner 100-200 $\mathrm{km}$ in radius. 
Here, fresh $^{23}$Na produced in carbon burning is mostly being converted via electron capture.
However, when considering the difference between the two simulations, FN2 and NB1, (the bottom plot of Fig.\ \ref{fig:electron}), we can see the additional changes to the electron fraction due to the convective Urca process. 
The inclusion of the convective Urca process in the FN2 simulation leads to a greater rate of electron captures near the center, and a greater rate of electrons deposited in the outer regions of the convection zone due to $\beta$-decays.

A negative electron fraction rate, as near the center, makes the material less buoyant. 
And vice-versa, a positive electron fraction rate, as around 550 km in the FN2 simulation, makes the fluid material more buoyant. 
In each case, the buoyancy of the fluid is subtly altered, which should alter the convective mixing. 
The magnitude and direction of these buoyancy shifts is less clear, but the magnitude should be proportional to the electron fraction rate, $\dot{\mathrm{Y}}_e(r)$.

The correlations between the Urca reactions, buoyancy, and subsequent impact on mixing is complex, and we have limited our exploration to a qualitative approach. 
But, the fact that the convective Urca process alters the buoyancy of the fluid may explain some of the differences in the convection zones observed between the simulations. 
We hope to explore this relation with a more quantitative approach in future work.

\subsection{Central Burning Evolution} \label{subsec:burn}
\begin{figure}[ht!]
\plotone{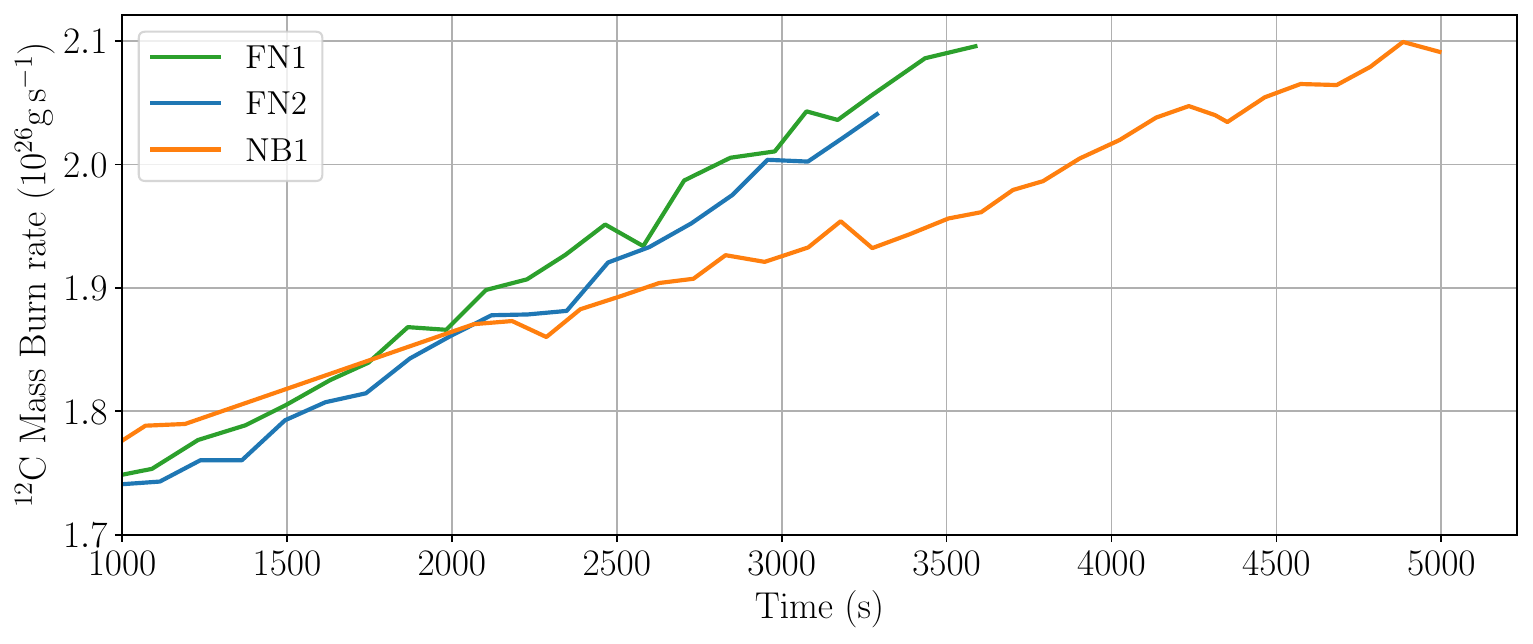}
\caption{The amount in grams of \isot{C}{12} burned in each simulation over the last few thousand seconds of simulation time. We plot all three simulations with FN1 in green, FN2 in blue and NB1 in orange.}
\label{fig:cburn_rate}
\end{figure}

The consequences of the convective Urca process and the difference in convection zone sizes likely leads to greater heating during the simmering phase.
Though the presented simulations only cover roughly an hour of simulation time, there is a noticeable difference in the rate of carbon burning at the end of the simulations, see Fig.\ \ref{fig:cburn_rate}.
Less efficient convection, via a smaller convection zone, appears to outweigh the neutrino losses due to the Urca reactions in this brief timespan.
Further work on earlier time periods is important and is discussed further in Section \ref{sec:7}. 

To summarize Section \ref{sec:5}, we present the most relevant quantitative results in Table \ref{tab:sum}.

\begin{table}[ht]
    \centering
    \begin{tabular}{c|c|c}
         & FN2 & NB1 \\
         \hline
         $R_\text{conv}$ & 543 km & 744 km\\
         \hline
         $M_\text{conv}$ & $0.66 \, \Msun$ & $1.0 \, \Msun$\\
         \hline
         $\dot{E}_\text{tot}$ & $5.44 \times 10^{43} \erg~\second^{-1}$ & $5.40 \times 10^{43} \erg~\second^{-1}$\\
         \hline
         $\dot{E}_\nu$ & ${-}4.25 \times 10^{42} \erg~\second^{-1}$ & ${-}3.61 \times 10^{42} \erg~\second^{-1}$\\
         \hline
         $E_\text{kin}$ & $7.27 \times 10^{44} \erg$ & $1.23 \times 10^{45} \erg$\\
         \hline
         $<U_\text{rms}>$ & $6 \kms$ & $7 \kms$\\
    \end{tabular}
    \caption{Summary of results}
    \label{tab:sum}
\end{table}

\section{Discussion}
\label{sec:6}
In the presented simulations of a WD in a late stage of the simmering phase, we found that the convective Urca process resulted in a ${\sim}30\%$ smaller convective core for the FN1/FN2 simulations compared to NB1. 
The resulting convection zones of FN1/FN2 still did extend significantly past the Urca shell ($\Rconvm \approx 545 \, \mathrm{km}$, $R_\mathrm{urca} = 412 \, \mathrm{km}$).
This result sits in contrast to the previous literature \citep{stein-wheeler2006} that predicted the convective core would be restricted to the Urca shell.
Note that this result does not imply that such restriction could never occur with different central conditions (i.e.\ earlier stages of the simmering phase), but it shows that the convective Urca process does not necessitate this restrictions to the Urca shell throughout the full simmering phase.

Considering the FN1 and FN2 simulations, we do not believe we successfully modeled the proper convective boundary. 
However, each simulation indicates convective boundaries and structures which we hope to explore in future work.
In the FN1 case, the extended semi-mixed region from about 550-725 km is likely a consequence of our initial model defining this region to be isentropic. 
Despite this, the final structure of the FN1 model may share some relation to a realistic split convection zone that could develop during the simmering phase. 
In this scenario, the inner convection zone would develop due to the usual carbon burning process, while the outer convection zone would form due to the energy deposition of $\beta$-decays from the Urca reactions.
The two zones would be connected by a semi-stable region (i.e.\ Ledoux stable but Schwarzschild unstable) similar to that seen in the green curves of Fig.\ \ref{fig:all_sims_conv_criteria}.
This structure of the convective regions of the white dwarf may only be temporary as the carbon burning will likely increase enough that the inner convection zone grows and engulfs the outer regions.

Informed by the FN1 simulation, the FN2 initial model was made to produce a convectively stable temperature gradient outside ${\sim}540 \, \mathrm{km}$.
The drawback of this model is we imposed a sub-adiabatic temperature gradient to connect the isentropic interior to the same outward temperature as the FN1/NB1 model.
This temperature gradient is unrealistic during the simmering phase, as the conduction timescales are too large for there to be substantial heating in convectively stable regions.
Instead, we propose that a more proper convective boundary would consist of a \textit{super}-adiabatic region that connects the interior isentropic convection zone to the exterior isothermal envelope.
This configuration could develop once a sufficiently large compositional gradient is produced by the carbon burning.
This compositional gradient, in addition to the convective Urca process, could restrict the growth of the convection zone. 
Carbon burning would still provide heating to the convection zone, raising the temperature throughout the zone, including at the convective boundary.
This in turn will cause a temperature discontinuity to form as the conductive timescales are long during the simmering phase.
To connect either side of this discontinuity requires a super-adiabatic temperature gradient that would be partially or fully stabilized by a compositional gradient.
A somewhat similar structure appears to have formed in the FN2 simulation (see Fig.\ \ref{fig:all_sims_conv_criteria}).
Further evidence of such a structure can be seen in 1D stellar evolutionary models \citep{piersanti2022}.

The role of the convective Urca process on this convective boundary is of significant interest. 
There are additional compositional gradients produced by the Urca reactions which can further stabilize this region. Additionally, as discussed in \ref{subsec:buoy}, the Urca reactions alter the buoyancy which may affect the convective mixing.
This is seen in the lower convective velocities between FN1/FN2 as compared to NB1, see Figures \ref{fig:mconv_vrms} and \ref{fig:urms}.
In the future, we hope to quantify the changes in buoyancy and their impact on the convective flow.

Lastly, concerning the effects of the energy losses produced due to the neutrino emission in the Urca reactions, we see that such losses are not enough to cool down the star. 
This result was already expected, since some previous literature commented that the neutrino losses were not enough to stop the thermonuclear runaway with the central conditions considered (B2025).
In fact, we see that these simulations suggest the convective Urca process accelerates the runaway, since the FN1 and FN2 simulations indicated larger increases in the carbon burning rate (see Fig.\ \ref{fig:cburn_rate}) compared to the NB1 simulation.
Note, the nuclear network used in these simulations is far from complete and this not a definitive result.
However, due to mixing of material far from the Urca shell, the Urca reactions tend to deposit more energy than is taken out from the neutrino emissions.
This, along with a restricted convection zone which is less efficient at distributing energy, likely leads to the convective Urca process accelerating the runaway/shortening the simmering phase.

An additional effect of a limited convection zone due to the convective Urca process is lowering the amount of carbon available to be burned. 
In our initial model, we assumed a uniform distribution of \Xisot{C}{12}.
However, the distribution of carbon at the start of simmering is more complex.
Near the center, there is less carbon due to prior burning during the AGB phase of the star (see \cite{piersanti2003} Fig.\ 1 and \cite{martinez-rodriguez2016} Fig.\ 4 for example distributions).
The exact distribution of carbon depends strongly on the prior stellar evolution and the $\isotm{C}{12}(\alpha, \gamma)\isotm{O}{16}$ rate. 
But, it is still clear from the 1D stellar models that convection zones of size $0.66 \, \Msun$ and $1 \, \Msun$, as seen here, can contain very different amounts of carbon.
In particular, a smaller convection zone will incorporate a smaller amount of carbon, lowering the \Xisot{C}{12} near the center where burning occurs.
This will likely change how the carbon burning progresses, as with all other conditions held constant, a smaller \Xisot{C}{12} will result in less carbon burning.
This may have a non-linear effect on the evolution of the WD and is best studied by considering specific cases in future 3D simulations, with additional context provided by 1D stellar evolution models.

We also note that the amount of Urca pair in our initial model ($\Xisotm{Na}{23}+\Xisotm{Ne}{23} = 5 \times 10^{-4}$) is relatively small.
This abundance was chosen to align with the expectations at the beginning of the simmering phase. 
However, \isot{Na}{23} is produced by carbon burning and thus in the later stages of simmering, which we simulate here, there should be a larger abundance of $A=23$ material.
This likely increases the effects of the convective Urca process, but to what extent is something we hope to study in future work.

Understanding the convective Urca process is important to capturing the proper evolution of the WD and its final composition just prior to the explosion. 
Shortening the simmering phase can result in less total carbon burning, and less neutronization. 
This effects some of the observational signatures of type Ia SNe, particularly the amount of $\isotm{Ni}{56}$ \citep{timmes2003}. 
Additionally, the size of the convective core during the simmering phase can influence the value of the central density of the star just prior to the explosion \citep{denissenkov2015}.

To conclude the discussion, we note that the results obtained about the effects of the convective Urca process are subjected to the stage of the simmering phase that we have chosen to simulate. Meaning that these results describe only the effects of the convective Urca process in a late stage of the simmering phase, but that may not be fully relatable at earlier stages. In the earlier stages of simmering, the carbon burning will be slower and as a result the convective timescales will be longer. This may allow the Urca reactions to have even more significant impact on the convection zone and evolution of the WD.

\section{Conclusions and Future Work}
\label{sec:7}
The convective Urca process plays a significant role in the simmering phase. Using \maestro\ we performed ``full-star" 3D hydrodynamic simulations of the convective zone of a WD in the simmering phase, following the work developed in \citet{willcox2018} and B2025. Through the comparison of these simulations we have been able to characterize the effects of the convective Urca process in a late stage of the simmering phase. 
The most relevant results obtained through such comparison include acknowledging that the convective Urca process limits the growth of the convective zone and slows down the convective mixing, which can result in an accelerated runaway of the WD. 
We have concluded that, even though the convective Urca process limits the extent of the convective region, this region can extend further than the Urca shell. 

In order to expand the results obtained in this work, further simulations will need to be performed. In particular, to make a better characterization of the convective Urca process effects, we are now starting to run simulations that include a more comprehensive nuclear network, accounting for additional carbon burning rates and other Urca pairs, as for example $\isotm{Ne}{21}-\isotm{F}{21}$ and $\isotm{Mg}{25}-\isotm{Na}{25}$ \citep{piersanti2022}. 
A large network will allow us to study the influence of multiple Urca shells and  better tracking the neutronization due to carbon burning. Furthermore, we are also interested in using different central conditions to investigate the impact of the convective Urca process on different stages of simmering. That includes central temperatures within the range $T_c \sim 1.5 - 4 \times 10^8 $ K. In addition, we will incorporate more realistic convective boundaries which reflect the influence of prior evolution.
We aim to produce a set of simulations that altogether can describe the longer-timescale evolution of the simmering phase, contributing to the development of a 1D stellar evolution model whose convective region parameters are fully informed by 3D hydrodynamic simulations.

\begin{acknowledgments}

The authors acknowledge the earlier dissertation work of Don Willcox \citep{willcox2018} on which this work builds.
\maestro\ is freely available on GitHub (\url{https://github.com/AMReX-Astro/}), and all problem setup files for the calculations presented here are in the code repository.  
This research was supported in part by the US Department of Energy (DOE) under grant DE-FG02-87ER40317.
This research has made use of NASA’s Astrophysics Data System Bibliographic Services.
The authors would like to thank Stony Brook Research Computing and Cyberinfrastructure, and the Institute for Advanced Computational Science at Stony Brook University for access to the high-performance SeaWulf computing system, which was made possible by a \$1.4M National Science Foundation grant (\#1531492). 
Lastly, we also acknowledge the CFIS Mobility Program for the partial funding of this
research work, particularly Fundaci\'{o} Privada Mir-Puig, CFIS partners, and donors of the crowdfunding program.
\end{acknowledgments}

\software{\amrex\ \citep{zhang2019},                 
          {\sffamily GNU Parallel} \citep{gnu_parallel},
          {\sffamily matplotlib}\  \citep{Hunter:2007},
          \maestro\ \citep{fan2019,maestroex_joss},
          {\sffamily NumPy}\ \citep{numpy2020},
          {\sffamily pandas} \citep{pandas},
          \pynucastro\ \citep{pynucastro,smith2023},
          {\sffamily SymPy} \citep{sympy}, 
          {\sffamily yt} \citep{turk2011}}
       
\appendix

\section{Thermal Properties of the Urca Reactions}
\label{sec:fermi}
Generally, when we refer to nuclear energy generated we are obtaining such values by considering the difference in rest mass between the reactants and the products, while accounting for the energy losses due to neutrino emission. 
This calculation gives the total energy that is injected into the star if the nuclear reaction that we consider does not involve electrons, as for example $\isotm{C}{12}$-burning reactions. 
However, in the case that electrons are involved, as in the Urca reactions, the Fermi energy associated with these electrons due to the degeneracy is considerably higher than its rest masses. Thus, careful consideration of the thermal properties of the Urca reactions is needed.

In particular, as was shown by \citet{bruenn1973} and studied in detail in \citet{barkat_wheeler1990}, an electron from the top of the Fermi sea in a degenerate star has an energy $\mu_{e} = \epsilon_F + m_{e}c^2 + k_{B} T$, where $\epsilon_F$ is the Fermi energy and $k_B T$ the kinetic contribution. However, due to the high degeneracy, we can assume $k_B T \ll \epsilon_F$, and the energy of an electron from the top of the Fermi sea can be written as $\mu_{e} \simeq \epsilon_F + m_{e}c^2$. Therefore, defining the excess thermal contribution, $\varepsilon_\text{e-cap}$, as the energy that is actually left behind in the medium in an electron capture reaction, we have:
\begin{equation}
\label{eq:b1}
    \varepsilon_\text{e-cap} = -\epsilon_{\text{th}} + \mu_{e} -\epsilon_\nu = \epsilon_\text{th} +\epsilon_F + m_{e}c^2 -\epsilon_\nu
\end{equation}
where $\epsilon_{\text{th}}$ represents the difference in rest mass between the nuclei ($\epsilon_\text{th}  = (m_{\isotm{Ne}{23}}-m_{\isotm{Na}{23}} - m_e)c^2 > 0$), and $\epsilon_\nu$ the energy lost due to neutrino emission. Note, we can consider that we are always capturing an electron from the top of the Fermi sea, because even if the electron captured has a lower energy level, it will create a hole in the Fermi sea that will be filled with an electron from a higher energy level, forming a cascade until there is a hole at the top energy level.

The same idea applies also for the $\beta$-decay reaction.
In this case, the amount of energy provided by the decay is distributed into the medium instead of the emerging electron itself, which should quickly settle down to near the top of the Fermi sea.
As a result we can write the energy left in the medium as: 
\begin{equation}
\label{eq:b2}
     \varepsilon_{\beta} = \epsilon_\text{th} - \mu_{e} -\epsilon_\nu = \epsilon_\text{th} -\epsilon_F - m_{e}c^2 -\epsilon_\nu
\end{equation}
Note, the Fermi energy is different from that considered in the electron capture reaction, since these reactions occur at regions with different densities. At the Urca shell $\mu_{e} = \epsilon_\text{th} $. So, at higher densities (smaller radii), $\mu_{e} > \epsilon_\text{th}$, and at lower densities (larger radii), $\mu_{e} < \epsilon_\text{th} $. Therefore, $\varepsilon_\text{e-cap}(\rho) > 0$ if $\mu_{e}(\rho) > \epsilon_\text{th} + \epsilon_\nu$, and that occurs for most densities larger than $\rho_\text{urca}$. Similar situation will occur for the $\beta$-decay reaction, and $\varepsilon_\beta(\rho) > 0$ for most densities smaller than $\rho_\text{urca}$. Meaning that in a context of degenerate matter, both Urca reactions contribute to heating the star if they occur far enough away from the Urca shell. Note that there is only a thin region around the Urca shell where the neutrino losses are producing a net cooling on a reaction-by-reaction basis. Only when $\rho \approx \rho_\text{urca}$, we have $\mu_{e}(\rho) \approx \epsilon_\text{th} $, and then $\varepsilon_\text{e-cap}(\rho) \approx -\epsilon_\nu < 0$ and $\varepsilon_\beta(\rho) \approx - \epsilon_\nu < 0$ will there be net cooling.

Therefore, when we split the energy contributions in three different components (nuclear energy, neutrino losses and compositional shift energy), we are basically breaking the sums in Equations \ref{eq:b1}, \ref{eq:b2} into their different terms. The compositional shifts contribution (or thermal contribution) is correlated with the different values of the Fermi energy at different densities. We express these terms as sums of the change in internal energy ($e$) with respect to the change in $X_k$ over all isotopes. In particular, the term that accounts for the compositional shifts energy ($\frac{\partial e}{\partial X_k}$) is defined as follows:
\begin{equation}
    \left.\frac{\partial e}{\partial X_k}\right|_{\rho, T, (X_j, j\neq k)}  = \left.\frac{\partial e }{\partial \bar{A}}\right|_{\rho, T, \bar{Z}}  \frac{\partial \bar{A} }{\partial X_k} + \left.\frac{\partial e}{\partial \bar{Z}}\right|_{\rho, T, \bar{A}}  \frac{\partial \bar{Z}}{\partial X_k}
\end{equation}
The EOS considers composition in terms of $\bar{A}$ and$\bar{Z}$, so we obtain values of $\frac{\partial e}{\partial X_k}$ by using the EOS. Thus, the actual formula that the EOS uses to compute $\frac{\partial e}{\partial X_k}$ is:
\begin{equation}
    \left.\frac{\partial e}{\partial X_k}\right|_{\rho, T, (X_j, j\neq k)} = \left.\frac{\partial e }{\partial \bar{A}}\right|_{\rho, T, \bar{Z}} \frac{\bar{A}}{A_k}  (A_k - \bar{A}) + \left.\frac{\partial e}{\partial \bar{Z}}\right|_{\rho, T, \bar{A}} \frac{\bar{A}}{A_k} (Z_k - \bar{Z})
\end{equation}
where $A_k$ and $Z_k$ are respectively the mass number and the atomic number of the k-th isotope, and $\bar{A}$ and $\bar{Z}$ are the averaged values in the cell.

\begin{figure}[ht!]
    \plotone{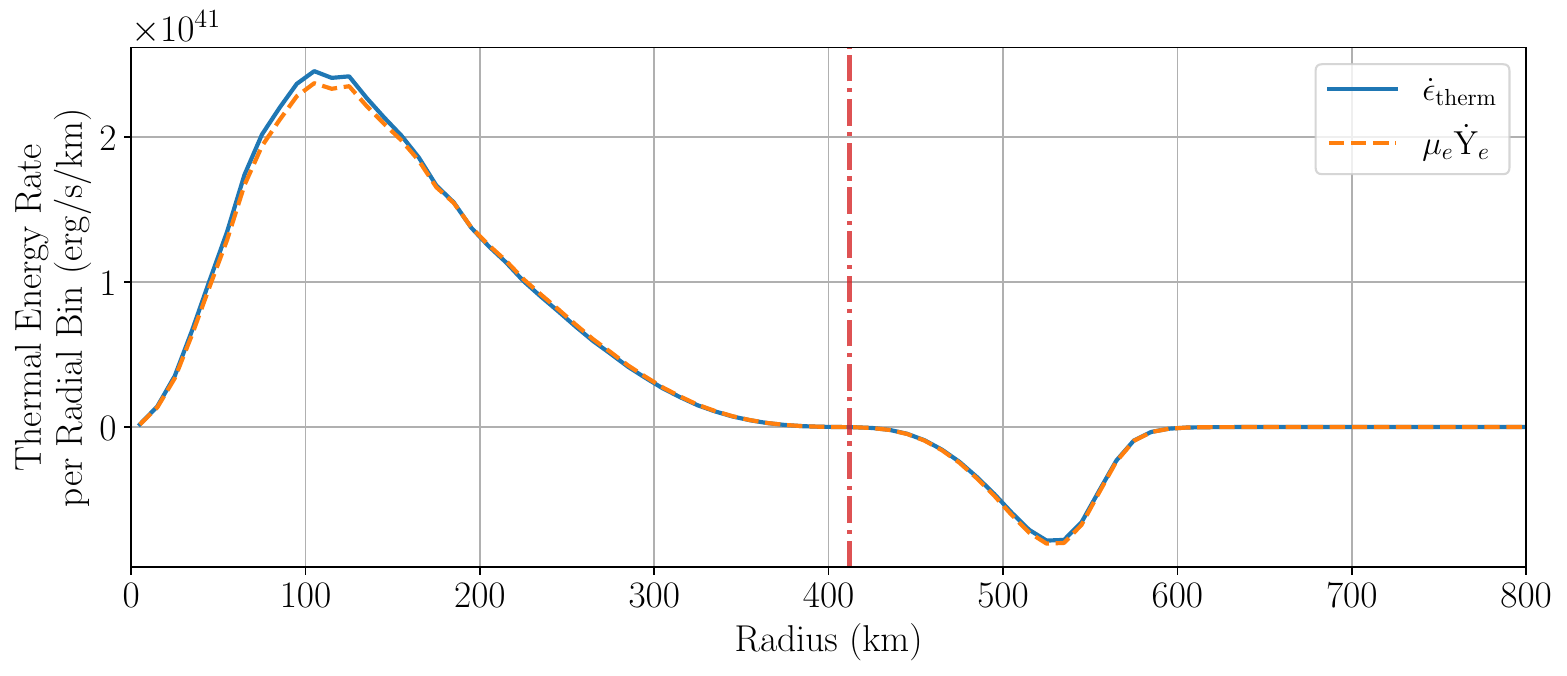}
    \caption{The thermal energy rate  per radial bin vs radius. Each curve is from the FN2 simulation. The blue curve includes all contributions to $\dot{\epsilon}_{\mathrm{therm}}$. The dashed orange curve includes the primary electron component. The red dash-dot vertical line indicates the location of the Urca shell.}
    \label{fig:therm_comp}
\end{figure}

Note this calculation includes contributions from ions, radiation and other reactions besides the contributions from the Urca reactions. However, the most significant portion does come from the electron component. We can isolate this component as,
\begin{equation}
    \epsilon_{e} = N_A \frac{d Y_e}{dt} \mu_e 
\end{equation}
where $N_A$ is Avogadro's number. We can then plot this individual component compared to the full thermal contributions due to compositional shifts in Fig.\ \ref{fig:therm_comp}. The thermal contributions are largely describing the contributions of electrons. 

\bibliography{urca}
\bibliographystyle{aasjournalv7}

\end{document}